\documentclass{article}

\usepackage[square,sort,comma,numbers,compress]{natbib}

\usepackage[final]{neurips_2021}

\usepackage[utf8]{inputenc} %
\usepackage[T1]{fontenc}    %
\usepackage{hyperref}       %

\usepackage{url}            %
\usepackage{booktabs}       %
\usepackage{amsfonts}       %
\usepackage{nicefrac}       %
\usepackage{microtype}      %
\usepackage{xcolor}         %
\usepackage{comment}
\usepackage{graphicx}
\usepackage{amsmath}
\usepackage{color}
\usepackage{appendix}
\usepackage{lipsum}
\usepackage{hanging}
\usepackage{wrapfig}
\usepackage{multirow}
\usepackage{enumitem}
\usepackage[skip=10pt]{subcaption}
\newcommand{\bb}[1]{\mathbf{#1}}

\newcommand{\bx}{\bb{x}}

\newcommand{\bg}{\bb{g}}
\newcommand{\bh}{\bb{h}}

\newcommand{\bz}{\bb{z}}

\newcommand{\bbf}{\bb{f}}

\newcommand{\bT}{\boldsymbol{\theta}}

\newcommand{\pT}{p_{\bT}}

\newcommand{\eg}{\textit{e}.\textit{g}., }

\title{PortaSpeech: Portable and High-Quality \\ Generative Text-to-Speech}
\author{
  Yi Ren\thanks{Equal contribution.}  \\
  Zhejiang University\\
  \texttt{rayeren@zju.edu.cn} \\
   \And
  Jinglin Liu\footnotemark[1] \\
  Zhejiang University\\
  \texttt{jinglinliu@zju.edu.cn} \\
  \And
  Zhou Zhao\thanks{Corresponding author} \\
  Zhejiang University \\
  \texttt{zhaozhou@zju.edu.cn} \\
}

\begin{document}
\maketitle
\begin{abstract}
Non-autoregressive text-to-speech (NAR-TTS) models such as FastSpeech 2~\citep{ren2020fastspeech} and Glow-TTS~\cite{kim2020glow} can synthesize high-quality speech from the given text in parallel. After analyzing two kinds of generative NAR-TTS models (VAE and normalizing flow), we find that: VAE is good at capturing the long-range semantics features (e.g., prosody) even with small model size but suffers from blurry and unnatural results; and normalizing flow is good at reconstructing the frequency bin-wise details but performs poorly when the number of model parameters is limited. Inspired by these observations, to generate diverse speech with natural details and rich prosody using a lightweight architecture, we propose PortaSpeech, a portable and high-quality generative text-to-speech model. Specifically, 1) to model both the prosody and mel-spectrogram details accurately, we adopt a lightweight VAE with an enhanced prior followed by a flow-based post-net with strong conditional inputs as the main architecture. 2) To further compress the model size and memory footprint, we introduce the grouped parameter sharing mechanism to the affine coupling layers in the post-net. 3) To improve the expressiveness of synthesized speech and reduce the dependency on accurate fine-grained alignment between text and speech, we propose a linguistic encoder with mixture alignment combining hard word-level alignment and soft phoneme-level alignment, which explicitly extracts word-level semantic information.  Experimental results show that PortaSpeech outperforms other TTS models in both voice quality and prosody modeling in terms of subjective and objective evaluation metrics, and shows only a slight performance degradation when reducing the model parameters to 6.7M (about 4x model size and 3x runtime memory compression ratio compared with FastSpeech 2). Our extensive ablation studies demonstrate that each design in PortaSpeech is effective\footnote{Audio samples are available at \url{https://portaspeech.github.io/}.}.

\end{abstract}

\section{Introduction}
\label{sec:intro}
Recently, deep learning-based text-to-speech (TTS) has attracted a lot of attention in speech community~\cite{wang2017tacotron,shen2018natural,ming2016deep,arik2017deep,ping2018deep,ren2019fastspeech,li2018close,ren2020fastspeech,liu2021diffsinger}. Among neural network-based TTS systems, some of them generate mel-spectrograms autoregressively from text~\cite{wang2017tacotron,shen2018natural,ping2018deep,li2018close} and suffer from slow inference speed and robustness (word skipping and repeating) problems~\cite{ren2019fastspeech}, while others~\cite{ren2019fastspeech,ren2020fastspeech,lancucki2020fastpitch,kim2020glow,lim2020jdi,miao2020flow} generate mel-spectrograms in parallel with comparable quality using non-autoregressive architecture, called NAR-TTS, which enjoys fast inference and avoids robustness issues in the meanwhile. In general, modern TTS models aim to achieve the following goals: 
\begin{itemize}[leftmargin=*]
    \item \textit{Fast}: to reduce the cost of computational resources and apply the model to real-time applications, the inference speed of TTS model should be fast.
    \item \textit{Lightweight}: to deploy the model to mobile or edge devices, the model size should be small and the runtime memory footprint should be low. 
    \item \textit{High-quality}: to improve the naturalness of synthesized speech, the model should capture the details (frequency bins between two adjacent harmonics, unvoiced frames and high-frequency parts) in natural speech.
    \item \textit{Expressive}: to generate expressive and dynamic speech, the model should use powerful prosody modeling methods to accurately model the fundamental frequency and duration of speech.
    \item \textit{Diverse}: to prevent the synthesized speech from being too dull and tedious when generating long speech, the model should be able to generate diverse speech samples with different intonations given one text input sequence.
\end{itemize}

To achieve the above goals, in this work, we propose PortaSpeech, a portable and high-quality generative text-to-speech model, which generates mel-spectrograms with natural details and expressive prosody using a lightweight architecture. Specifically,
\begin{itemize}[leftmargin=*]
    \item Through some preliminary experiments (see \autoref{sec:ana_vae_flow}), we find that VAE is good at capturing the long-range semantics features (\eg prosody), while normalizing flow is good at reconstructing the frequency bin-wise details. Based on these observations, we adopt VAE with an enhanced prior followed by a flow-based post-net as the main model architecture of PortaSpeech, which helps PortaSpeech generate \textit{high-quality} and \textit{expressive} results. In addition, PortaSpeech can generate \textit{diverse} speech by sampling latent variables from the prior of VAE and post-net.
    \item Through the experiments, we also find that even when the model is very small, VAE is still good at capturing the prosody, making it possible for PortaSpeech to reduce its model size using a lightweight VAE. Besides, we introduce the grouped parameter sharing mechanism to the post-net to compress its model size. By doing these, PortaSpeech can be very \textit{lightweight} and \textit{fast} at a small performance cost.
    \item To model the prosody better and generate more \textit{expressive} speech, we introduce a linguistic encoder with mixture alignment, which combines hard word-level alignment and soft phoneme-level alignment. Our proposed linguistic encoder also reduces the dependence on fine-grained (phoneme-level) alignment and alleviates the burden of the speech-to-text aligner.
\end{itemize}

Experiments on the LJSpeech~\citep{ljspeech17} dataset show that PortaSpeech outperforms other state-of-the-art TTS models with comparable model parameters in voice quality and prosody in terms of both subjective and objective evaluation metrics. When compressing the model size, our PortaSpeech shows only a slight performance degradation but enjoys the benefits of a much smaller number of model parameters (about 4x model size reduction) and lower memory footprints (about 3x memory reduction) compared with FastSpeech 2. The main contributions of this work are summarized as follows:
\begin{itemize}[leftmargin=*]
\item We analyze the characteristics of VAE and normalizing flow when applied to TTS and combines the advantages of VAE and normalizing flow to generate mel-spectrograms with rich details and expressive prosody.
\item We propose mixture alignment in the linguistic encoder, which improves the prosody and reduces the dependence on fine-grained (phoneme-level) hard alignment.
\item Using lightweight VAE and introducing the grouped parameter sharing mechanism to the post-net, PortaSpeech can generate high-quality speech with a small number of model parameters and small runtime memory footprints.
\end{itemize}

\section{Background}
In this section, we describe the background of TTS and the basic knowledge of VAE and normalizing flow. We also review the existing applications of VAE and normalizing flow in non-autoregressive TTS and analyze their advantages and disadvantages.

\paragraph{Text-to-Speech}
Text-to-speech (TTS) models convert input text or phoneme sequence into mel-spectrogram (\eg Tacotron~\cite{wang2017tacotron}, FastSpeech~\cite{ren2019fastspeech}), which is then transformed to waveform using vocoder (\eg WaveNet~\cite{van2016wavenet}), or directly generate waveform from text (\eg FastSpeech 2s~\cite{ren2020fastspeech} and EATS~\cite{donahue2020end}). End-to-end text-to-speech models have gradually developed from autoregressive to non-autoregressive architecture: early autoregressive text-to-speech models~\cite{wang2017tacotron,shen2018natural} generate each mel-spectrogram frame conditioned on previous ones, resulting in high inference latency and low robustness. Recently, several non-autoregressive TTS works have been proposed, which generate mel-spectrogram frames in parallel. FastSpeech~\cite{ren2019fastspeech} and ParaNet~\cite{peng2020non} are the first non-autoregressive TTS models, which use pre-trained autoregressive TTS teacher models to extract text-to-spectrogram alignments from the training data to bridge the length gap between text and speech for non-autoregressive student model. FastSpeech 2~\cite{ren2020fastspeech} introduces more variation information of speech, including pitch and energy, to alleviate the one-to-many mapping problem in TTS. While these methods need external text-to-spectrogram alignment models or tools, Glow-TTS~\cite{kim2020glow} directly searches for the most probable monotonic alignment between text and the latent representation of speech using normalizing flows and dynamic programming. In addition to improving the performance of non-autoregressive models, some works focus on lightweight and portable model designs: SpeedySpeech~\cite{vainer2020speedyspeech} replaces the self-attention layers with fully convolutional blocks to reduce the computational complexity. LightSpeech~\cite{luo2021lightspeech} leverages neural architecture search (NAS) to automatically design more lightweight models, while the training of NAS consumes huge resources. In this work, we save the model parameters by taking advantage of the characteristics of VAE and normalizing flow and introducing the grouped parameter sharing mechanism.

\paragraph{VAE} 
\label{sec:bg_vae}
The VAE is a generative model in the form of $p_{\theta}(\bx,\bz)=p(\bz)p_{\theta}(\bx|\bz)$, where $p(\bz)$ is a prior distribution over latent variables $\bz$ and $p_{\theta}(\bx|\bz)$ is the likelihood function that generates data $x$ given latent variables $\bz$ which can be considered as a decoder. It is parameterized by a neural network $\theta$. Since the true posterior $p_{\theta}(\bx, \bz)$ over the latent variables of a VAE is usually analytically intractable, we approximate it with a variational distribution $q_{\phi}(\bz|\bx)$, which can be viewed as an encoder. The parameters $\theta$ and $\phi$ can be optimized by maximizing the \textit{evidence lower bound} (ELBO):
\begin{align}
    \log p_{\theta}(\bx) \ge& \mathbb{E}_{q_{\phi}(\bz|\bx)}\left[\log \frac{p_{\theta}(\bx,\bz)}{q_{\phi}(\mathbf{\bz}|\bx)} \right]
=E_{\bz \sim q_{\phi}(\bz|\bx)} \left[\log p_{\theta}(\bx|\bz)- \log \frac{q_{\phi}(\bz|\bx)}{p_{\theta}(\bz)}\right] \nonumber \\ 
=&E_{\bz \sim q_{\phi}(\bz|\bx)} \left[\log p_{\theta}(\bx|\bz)\right]
-\operatorname{KL}\left(q_{\phi}(\bz|\bx) \| p_{\theta}(\bz)\right)
    \equiv \mathcal{L}(\theta, \phi). \nonumber
\end{align}
Recently, some works successfully apply VAE to TTS. One of them is BVAE-TTS~\cite{lee2020bidirectional}, which adopts a bidirectional-inference variational autoencoder that learns hierarchical latent representations using both bottom-up and top-down paths to increase its expressiveness. Thanks to the hierarchical structure and latent modeling, BVAE-TTS can capture the dynamism and variability of ground-truth prosody. However, its generated mel-spectrograms are very blurry and over-smoothing, resulting in unnatural sounds, due to the posterior collapse~\cite{alemi2018fixing,he2019lagging} and the reconstruction loss term used in BVAE-TTS, which has independency assumption of generated frequency bins given latent variables.

\paragraph{Normalizing Flow}
\label{sec:bg_flow}
Normalizing flow is a kind of generative models~\citep{dinh2014nice,dinh2016density} which has several advantages including exact log-likelihood evaluation and fully-parallel sampling. In generation, normalizing flows~\citep{dinh2014nice,dinh2016density} transform the latent variable $\bz$ into a datapoint $\bx$ through a composition of invertible functions $\bbf = \bbf_1 \circ \bbf_2 \circ \cdots \circ \bbf_K$ and we assume a tractable prior $\pT(\bz)$ over latent variable $\bz$ sampled from a simple distribution (\eg a Gaussian distribution). In training, the log-likelihood of a datapoint $\bx$ can be computed exactly using the change of variables rule:
\begin{align}
\label{eq:loglike2}
\log \pT(\bx) &= \log \pT(\bz) + \sum_{i=1}^{K} \log|\det(d\bh_i/d\bh_{i-1})|,
\end{align}
where $\bh_0 = \bx$, $\bh_i = \bbf_{i}(\bh_{i-1})$, $\bh_K = \bz$ and $|\det(d\bh_i/d\bh_{i-1})|$ is the Jacobian determinant. We learn the parameters of $ \bbf_1 \dots \bbf_K $ by maximizing Equation \eqref{eq:loglike2} over the training data. Given $\bg = \bbf^{-1}$, we can now generate a sample $\hat{\bx}$ by sampling $ \bz \sim \pT(\bz)$ and computing $\hat{\bx} = \bg(\bz)$.

There are several normalizing flow-based non-autoregressive TTS methods: Flow-TTS~\cite{miao2020flow} is an early flow-based TTS method, which replaces the decoder in FastSpeech with Glow~\cite{kingma2018glow} and jointly learns the alignment and mel-spectrogram generation through a single network. Then Glow-TTS~\cite{kim2020glow} is proposed, which combines the normalizing flow and dynamic programming-based monotonic alignment to enable fast, diverse and controllable speech synthesis. These methods handle the blurry mel-spectrogram problems well due to the nature of the normalizing flow. However, according to our experiments (see \autoref{sec:ana_vae_flow}), flow-based NAR-TTS model usually requires a huge model capacity to achieve good performance, and the performance can drop notably when reducing the number of model parameters.

\section{PortaSpeech}
\begin{figure}[!t]
	\centering
	\begin{subfigure}[h]{0.24\textwidth}
		\centering
		\includegraphics[width=\textwidth,trim={0cm 0.05cm 9.2cm 0cm}, clip=true]{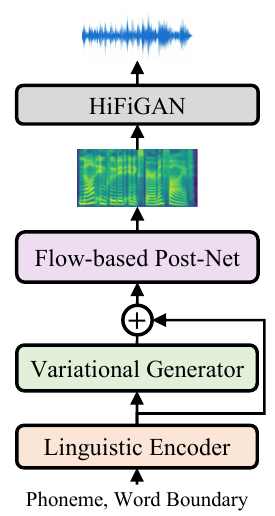}
		\caption{PortaSpeech}
		\label{fig:arch_1}
	\end{subfigure}
	\begin{subfigure}[h]{0.24\textwidth}
		\centering
		\includegraphics[width=\textwidth,trim={2.9cm 0.05cm 6.3cm 0cm}, clip=true]{figs/arch.pdf}
		\caption{Linguistic Encoder}
		\label{fig:arch_2}
	\end{subfigure}
	\begin{subfigure}[h]{0.22\textwidth}
		\centering
		\includegraphics[width=\textwidth,trim={5.7cm 0.33cm 3.9cm 0.07cm}, clip=true]{figs/arch.pdf}
		\caption{Variational Generator}
		\label{fig:arch_3}
	\end{subfigure}
	\begin{subfigure}[h]{0.26\textwidth}
		\centering
		\includegraphics[width=\textwidth,trim={8.2cm 0.17cm 0.6cm 0.0cm}, clip=true]{figs/arch.pdf}
		\vspace{0.5mm}
		\caption{Flow-based Post-Net}
		\label{fig:arch_4}
	\end{subfigure}
	\caption{The overall architecture for PortaSpeech. In subfigure (b), "WP" denotes the word-level pooling operation, "LR" denotes the length regulator proposed in FastSpeech and "sinusoidal-like symbol" denotes the positional encoding. In subfigure (c), "VP-Flow" denotes the volume-preserving normalizing flow. In subfigure (c) and (d), the operations denoted with dotted lines are only used in the training procedure.}
	\label{fig:arch}
\end{figure}

Considering the characteristics of VAE and normalizing flow mentioned in \autoref{sec:bg_vae}, to build a TTS system that can meet the goals described in \autoref{sec:intro}, we propose PortaSpeech, which combines the advantages of VAE and normalizing flows and overcomes their deficiencies. 
As shown in \autoref{fig:arch_1}, PortaSpeech is composed of a linguistic encoder with mixture alignment, a variational generator with enhanced prior and a flow-based post-net with the grouped parameter sharing mechanism. First, the text sequence with word-level boundary is fed into the linguistic encoder to extract the linguistic features in both phoneme and word level. 
Secondly, to model the expressiveness and variability of speech with lightweight architecture, we train the VAE-based variational generator to maximize the ELBO over the ground-truth mel-spectrograms conditioned on the linguistic features, whose prior distribution is modeled by a small volume-preserving normalizing flow. Finally, to refine and enhance the natural speech details in the generated mel-spectrograms, we train the post-net by maximizing the likelihood of ground-truth mel-spectrograms conditioned on both the linguistic features and the outputs of the variational generator. During inference, the text is transformed to mel-spectrograms by successively passing through the linguistic encoder, the decoder of the variational generator and the reversed flow-based post-net. We describe these designs and the training and inference procedures in detail in the following subsections. We put more details in Appendix \ref{apdx:model_details}.

\subsection{Linguistic Encoder with Mixture Alignment}
To expand the lengths of linguistic features (outputs of the linguistic encoder), previous non-autoregressive TTS models introduce a duration predictor to predict the number of frames of each phoneme (phoneme duration) and the ground-truth phoneme duration (hard alignment) is obtained by external models/tools (\eg FastSpeech~\cite{ren2019fastspeech} and FastSpeech 2~\cite{ren2020fastspeech}) or jointly monotonic alignment training (\eg Glow-TTS~\cite{kim2020glow} and BVAE-TTS~\cite{lee2020bidirectional}). However, phoneme-level hard alignment has several issues: since some of the boundaries between two phonemes are naturally uncertain\footnote{It could be difficult to determine the exact boundary between two phonemes in millisecond level even for manually labeling.}, it is challenging for the alignment model to obtain very accurate phoneme-level boundaries, which inevitably introduces errors and noises. Further, these alignment errors and noises can affect the training of duration predictor, which hurts the prosody of the generated speech in inference. To tackle these problems, we introduce mixture alignment to the linguistic encoder, which uses soft alignment in phoneme level and keeps hard alignment in word level. 

As shown in \autoref{fig:arch_2}, our linguistic encoder consists of a phoneme encoder, a word encoder, a duration predictor and a word-to-phoneme attention module and detailed architecture of these modules are put in Appendix \ref{apdx:ling_enc}. Suppose we have an input phoneme sequence together with the word boundary (for example, "HH AE1 Z | N EH1 V ER0", where "|" denotes the word boundary in phoneme sequence). First, we encode the phoneme sequence into phoneme hidden states $\mathcal{H}_p$. Then we apply word-level pooling on $\mathcal{H}_p$ to obtain the input representation of the word encoder, which averages the phoneme hidden states inside each word according to the word boundary. The word encoder then encodes the word-level hidden states into word-level hidden states and expanded them to match the length of the target mel-spectrogram (denoted as $\mathcal{H}_w$) using length regulator with the word-level duration. Finally, to add fine-grained linguistic information, we introduce a word-to-phoneme attention module, which takes $\mathcal{H}_w$ as the query and $\mathcal{H}_p$ as the key and the value. In addition, due to the monotonic nature of text-to-spectrogram alignment, to encourage the attention to be close to the diagonal, we add a word-level relative positional encoding embedding to both $\mathcal{H}_p$ and $\mathcal{H}_w$ before they are fed into the attention module. To predict the word-level duration, we use the duration predictor which takes $\mathcal{H}_p$ as input and then sums the predicted duration of the phonemes in each word as the word-level duration\footnote{In training, the ground-truth word-level duration can be obtained by external forced alignment tools or autoregressive TTS models.}. Our mixture alignment mechanism avoids the uncertain and noisy phoneme-level alignment extraction and duration prediction while keeping fine-grained, soft and close-to-diagonal text-to-spectrogram alignment. 

\subsection{Variational Generator with Enhanced Prior}
To achieve \textit{expressive} and \textit{diverse} speech generation with \textit{lightweight} architecture, we introduce VAE as the mel-spectrogram generator, called variational generator. However, traditional VAE uses simple distribution (\eg Gaussian distribution) as the prior, which results in strong constraints on the posterior: optimizing with Gaussian prior pushes the posterior distribution towards the mean, limiting diversity and hurting the generative power~\cite{mahajan2020latent,tomczak2018vae}. To enhance the prior distribution, inspired by \cite{mahajan2020latent,setiawan2020variational,rezende2015variational,kingma2016improved}, we introduce a small volume-preserving normalizing flow\footnote{For simplicity and convenience, we use volume-preserving flow (VP-Flow), which does not need to consider the Jacobian term when calculating the data log-likelihood. We find that volume-preserving is powerful enough for modeling the prior.}, which transforms simple distributions (\eg Gaussian distribution) to complex distributions through a series of K invertible mappings (a stack of WaveNet residual blocks with dilation 1). Then we take the complex distributions as the prior of the VAE. When introducing normalizing flow-based enhanced prior, the optimization objective of the mel-spectrogram generator becomes: \begin{equation}
\log p(\bx|c)\geq\mathbb{E}_{q_\phi(\bz|\bx, c)}[\log p_\theta(\bx|\bz, c)]-\operatorname{KL}(q_{\phi}(\bz|\bx, c)|p_{\bar{\theta}}(\bz|c)) \equiv \mathcal{L}(\phi, \theta, \bar{\theta}),\label{eq:flowvae_px}
\end{equation}
where $\phi$, $\theta$ and $\bar{\theta}$ denote the model parameters of VAE encoder, VAE decoder and the normalizing flow-based enhanced prior, respectively; $c$ denotes the outputs of linguistic encoder. Due to the introduction of normalizing flows, the KL term in Equation \eqref{eq:flowvae_px} no longer offers a simple closed-form solution. So we estimate the expectation w.r.t. $q_{\phi}(\bz|\bx,c)$ via Monte-Carlo method by modifying the KL term: 
\begin{equation}
\operatorname{KL}(q_{\phi}(\bz|\bx, c)|p_{\bar{\theta}}(\bz|c)) = \mathbb{E}_{q_\phi(\bz|\bx,c)}[\log q_{\phi}(\bz|\bx,c) - \log p_{\bar{\theta}}(\bz|c)].
\label{eq:vae_kl}
\end{equation}

As shown in \autoref{fig:arch_3}, in training, the posterior distribution $N(\mu_q, \sigma_q)$ is encoded by the encoder of the variational generator. Then $z_q$ is sampled from the posterior distribution using reparameterization and is passed to the decoder of the variational generator (the right dotted line). In the meanwhile, the posterior distribution is fed into the VP-Flow to convert it to a standard normal distribution (the middle dotted line). In inference, VP-Flow converts a sample in the standard normal distribution into a sample $z_p$ in the prior distribution of the variational generator and we pass the $z_p$ to the decoder of the variational generator.

\subsection{Flow-based Post-Net}
To generate \textit{high-quality} mel-spectrograms, normalizing flows~\cite{kim2020glow,miao2020flow} have been widely proved to be effective. Unlike simple loss-based (L1 or MSE-based) or VAE-based methods that often generate blurry outputs, flow-based models can overcome the over-smoothing problem and generate more realistic outputs. To model rich details in ground-truth mel-spectrograms, we introduce a flow-based post-net with strong condition inputs to refine the outputs of the variational generator. As shown in \autoref{fig:arch_4}, the architecture of the post-net adopts Glow~\cite{kingma2018glow} and is conditioned on the outputs of the variational generator and the linguistic encoder. In training, the post-net transforms the mel-spectrogram samples into latent prior distribution (isotropic multivariate Gaussian) and calculates the exact log-likelihood of the data using the change of variables. In inference, we sample the latent variables from the latent prior distribution and pass them into the post-net reversely to generate the high-quality mel-spectrogram.

\begin{wrapfigure}{r}{4cm}
    \vspace{-2mm}
    \centering
	\includegraphics[width=0.3\textwidth,trim={0cm 0cm 0cm 0cm}, clip=true]{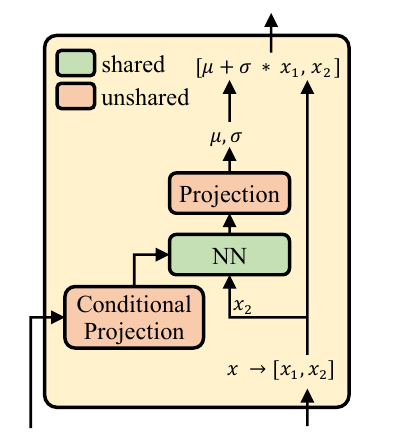}
	\caption{Affine coupling layer with grouped parameter sharing. Green block means sharing the model parameters of this block among flow layers in a group.}
	\label{fig:arch_5}
	\vspace{-3mm}
\end{wrapfigure}

However, flow-based models suffer from large model footprints. Since the conditional inputs contain the text and prosody information, our post-net only focuses on modeling the details in mel-spectrograms, greatly reducing requirements for model capacity. To further reduce the model size and keep the modeling power, we introduce the grouped parameter sharing mechanism to the affine coupling layer, which shares some model parameters among different flow steps ($\bbf_i$, $\bbf_{i+1}$, ..., $\bbf_j$). As shown in \autoref{fig:arch_5}, we divide all flow steps ($\bbf_1$, $\bbf_2$, ..., $\bbf_K$) into several groups and share the model parameters of $NN$ (WaveNet-like network, see Appendix \ref{apdx:pn}) in the coupling layers among flow steps in a group. Our grouped parameter sharing mechanism is similar to the shared neural density estimator proposed in \cite{lee2020nanoflow} with some differences that: 1) we simplify the model by removing the flow indication embedding since the unshared conditional projection layer in different flow steps can help the model to indicate the position of the step; 2) instead of sharing the parameters among all flow steps, we generalize the sharing mechanism by sharing the parameters among flow steps in a group, making it easier to adjust the number of trainable model parameters without changing the model architecture.

\subsection{Training and Inference}
In training, the final loss of PortaSpeech consists of the following loss terms: 1) duration prediction loss $L_{dur}$: MSE between the predicted and the ground-truth word-level duration in log scale; 2) reconstruction loss of variational generator $L_{VG}$: MAE between the ground-truth mel-spectogram and that generated by the variational generator; 3) the KL-divergence of variational generator $L_{KL}$ = $\log q_{\phi}(\bz|\bx,c) - \log p_{\bar{\theta}}(\bz|c)$, where $\bz \sim q_\phi(\bz|\bx,c)$, according to Equation \eqref{eq:vae_kl}; and 4) the negative log-likelihood of the post-net $L_{PN}$. In inference, the linguistic encoder first encodes the text sequence, predicts the word-level duration and expand the hidden states via mixture alignment to obtain the linguistic hidden states $\mathcal{H}_L$. Secondly, we sample $\bz$ from the enhanced prior, and then the decoder of the variational generator generates the coarse-grained mel-spectrograms $\bar{M}_c$ (the output mel-spectrograms before post-net) conditioned on the linguistic hidden states $\mathcal{H}_L$. Thirdly, the post-net converts randomly sampled latent into fine-grained mel-spectrograms $\bar{M}_f$ conditioned on $\mathcal{H}_L$ and $\bar{M}_c$. Finally, $\bar{M}_f$ is transformed to waveform using a pre-trained vocoder. Since we use hard word-level alignment in PortaSpeech, absolute durations for individual words can also be specified at inference time like FastSpeech~\cite{ren2019fastspeech}. As for silences, we add a word boundary symbol as an extra special word such as "SIL" between two words in training. In this way, we can adjust the duration of silences via modifying the duration of the special word "SIL".

\section{Experiments}
\subsection{Experimental Setup} 
\label{sec:exp_setting}
\paragraph{Datasets} 

We evaluate PortaSpeech on LJSpeech dataset~\citep{ljspeech17}, which contains 13100 English audio clips and corresponding text transcripts. Following FastSpeech 2~\cite{ren2020fastspeech}, we split LJSpeech dataset into three subsets: 12229 samples for training, 348 samples (with document title LJ003) for validation and 523 samples (with document title LJ001 and LJ002) for testing. We randomly choose 50 samples in the test set for subjective evaluation and use all testing samples for objective evaluation. We convert the text sequence to the phoneme sequence~\citep{arik2017deep,wang2017tacotron,shen2018natural,sun2019token,ren2019fastspeech} with an open-source grapheme-to-phoneme tool\footnote{\url{https://github.com/Kyubyong/g2p}}. We transform the raw waveform with the sampling rate 22050 into mel-spectrograms following \cite{shen2018natural,ren2019fastspeech} with the frame size 1024 and the hop size 256.

\paragraph{Model Configuration}
Our PortaSpeech consists of an encoder, a variational generator and a post-net. The encoder consists of multiple feed-forward Transformer blocks~\citep{ren2019fastspeech} with relative position encoding~\cite{shaw2018self} following Glow-TTS~\cite{kim2020glow}. The encoder and decoder in variational generator are 2D-convolution networks. The post-net adopts the architecture of Glow~\cite{kingma2018glow}. We conduct experiments on two settings with different model sizes: \textit{PortaSpeech (normal)} and \textit{PortaSpeech (small)}. We add more detailed model configurations of these two settings in Appendix \ref{apdx:exp_settings}.

\paragraph{Training and Evaluation}
\label{sec:exp_train_eval}
We train the PortaSpeech on 1 NVIDIA 2080Ti GPU, with batch size of 64 sentences on each GPU. We use the Adam optimizer with $\beta_{1}= 0.9$, $\beta_{2} = 0.98$, $\varepsilon = 10^{-9}$ and follow the same learning rate schedule in \citep{vaswani2017attention}. It takes 320k steps for training until convergence. 
The output mel-spectrograms of our model are transformed into audio samples using HiFi-GAN~\citep{kong2020hifi}\footnote{\url{https://github.com/jik876/hifi-gan}} trained in advance. We conduct the MOS (mean opinion score) and CMOS (comparative mean opinion score) evaluation on the test set to measure the audio quality via Amazon Mechanical Turk. We keep the text content consistent among different models to exclude other interference factors, only examining the audio quality or prosody. Each audio is listened by at least 20 testers. We analyze the MOS and CMOS in two aspects: prosody (naturalness of pitch, energy and duration) and audio quality (clarity, high-frequency and original timbre reconstruction), and score MOS-P/CMOS-P and MOS-Q/CMOS-Q corresponding to the MOS/CMOS of prosody and audio quality. We tell the tester to focus on one aspect and ignore the other aspect when scoring MOS/CMOS of this aspect. We put more information about the subjective evaluation in Appendix \ref{apdx:suj_eval_details}.

\subsection{Preliminary Analyses on VAE and Flow}
\label{sec:ana_vae_flow}

In image generation tasks, VAE is good at capturing the overall image structure information (low-frequency parts) while discarding small sharp textures/details (high-frequency parts). Similarly, in mel-spectrograms, low-frequency parts correspond to the shape of harmonics, which determines the pitch and prosody of speech. Thus we can intuitively infer that VAE is good at modeling the prosody while not good at modeling the details in speech. While flow-based models can generate high-quality images at the cost of very large model size and huge computation complexity and we may infer that flow-based models can model the details in speech well with large model size.

\begin{table}[!h]
\small
\vspace{-2mm}
\caption{The audio performance comparisons among different NAR-TTS models with different numbers of model parameters (\#Params.). GT (voc.) denotes the waveform reconstructed from ground-truth mel-spectrograms using HiFi-GAN~\cite{kong2020hifi}.}
\label{tab:exp_model_size}
\vspace{2mm}
\centering
\begin{tabular}{ c | c | c | c | c}
\toprule
Methods                                     & Configs & MOS-P           & MOS-Q           & \#Params \\
\midrule
\textit{GT (voc.)}                          & /      & 4.49 $\pm$ 0.07  & 4.16 $\pm$ 0.06 & /     \\
\midrule  
\multirow{3}{*}{\textit{Flow-based}}        & big    & 3.71 $\pm$ 0.06  & \textbf{3.96 $\pm$ 0.07}  & 41.2M \\
                                            & middle & 3.52 $\pm$ 0.07  & 3.54 $\pm$ 0.12  & 10.2M \\
                                            & small  & 3.21 $\pm$ 0.12  & 3.42 $\pm$ 0.14  & 4.5M \\
\midrule  
\multirow{3}{*}{\textit{VAE-based}}         & big    & \textbf{3.81 $\pm$ 0.07}  & 3.75 $\pm$ 0.08 & 43.2M  \\
                                            & middle & 3.79 $\pm$ 0.08  & 3.69 $\pm$ 0.09 & 9.3M \\
                                            & small  & 3.72 $\pm$ 0.08  & 3.51 $\pm$ 0.11 & 4.4M \\
\bottomrule
\end{tabular}
\end{table}

To verify our hypothesis and explore the characteristic of VAE and flow-based models in TTS, we conduct audio quality (MOS-Q) and prosody (MOS-P) comparisons among several VAE and flow-based NAR-TTS models with different model sizes: 1) \textit{big}: more than 40M model parameters; 2) \textit{middle}: about 10M model parameters; and 3) \textit{small}: about 5M model parameters. We keep the architecture of the encoders in three models consistent. The detailed model architecture and configurations are put in Appendix \ref{apdx:model_ana}. The results are shown in \autoref{tab:exp_model_size}. From the table, we can see that 1) when reducing the model capacities, the prosody quality of flow-based models drops significantly. In contrast, that of VAE-based model only drops slightly, according to MOS-P. This phenomenon inspires us to apply VAE-based mel-spectrogram decoder (variational generator) to our lightweight TTS model. 2) Compared with flow-based models, VAE-based model has poorer audio quality upper bound according to MOS-Q, which motivates us to make up for shortcomings of VAE by introducing a flow-based post-net to refine the mel-spectrograms generated by VAE.

\subsection{Performance}
\newcommand{\plotmelL}[2]{
    \begin{subfigure}{0.235\textwidth}
	\centering
	\includegraphics[width=\textwidth,trim={2mm 2mm 2mm 2mm}, clip=true]{#1}
	\vspace{-2mm}
	\caption{\textit{#2}}
	\vspace{2mm}
    \end{subfigure}
}
\begin{figure}[!t]
    \plotmelL{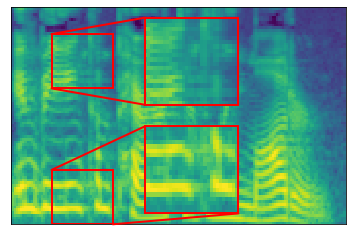}{GT}
    \plotmelL{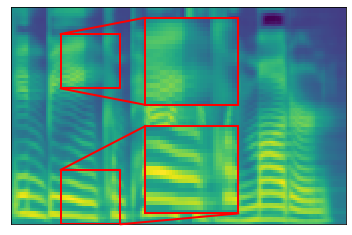}{TransformerTTS}
    \plotmelL{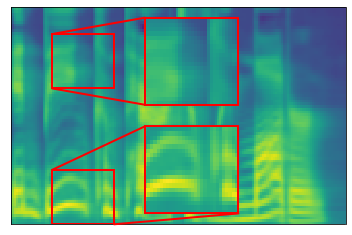}{FastSpeech}
    \plotmelL{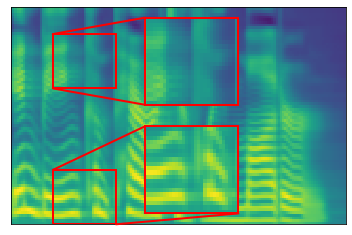}{FastSpeech 2}
    \plotmelL{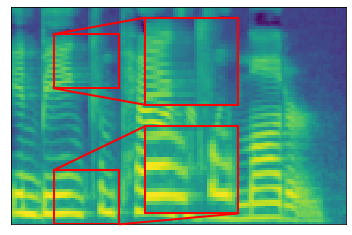}{Glow-TTS}
    \plotmelL{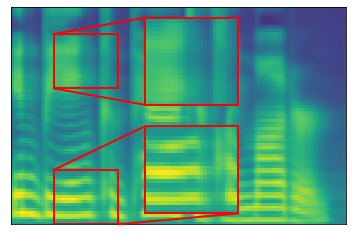}{BVAE-TTS}
    \plotmelL{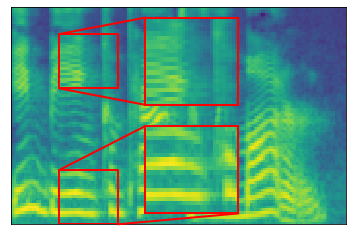}{PortaSpeech (normal)}
    \plotmelL{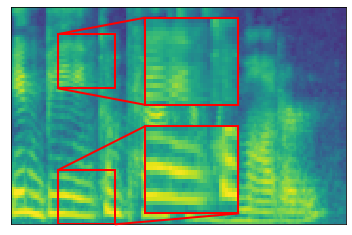}{PortaSpeech (small)}
    \caption{Visualizations of the ground-truth and generated mel-spectrograms by different TTS models. The corresponding text is "\textit{In being comparatively modern}".}
    \label{fig:mel_main_vis}
\end{figure}

\begin{table}[!h]
\small
\vspace{-2mm}
\caption{The audio performance (MOS-Q and MOS-P), inference latency, peak memory (Peak Mem.) and number of model parameters (\#Params.) comparisons. The evaluation is conducted on a server with 1 NVIDIA 2080Ti GPU and batch size 1. The mel-spectrograms are converted to waveforms using Hifi-GAN (V1)~\cite{kong2020hifi}. RTF denotes the real-time factor, that the seconds required for the system (together with Hifi-GAN vocoder) to synthesize one-second audio.}
\vspace{2mm}
\centering
\begin{tabular}{ l | c | c | c | c | c }
\toprule
Method &  MOS-P & MOS-Q & RTF & Peak Mem. & \#Params. \\
\midrule
\textit{GT}                                    & 4.52 $\pm$ 0.07 & 4.41 $\pm$ 0.06 & /    & /       & /   \\
\textit{GT (voc.)}                             & 4.48 $\pm$ 0.08 & 4.15 $\pm$ 0.07 & /    & /       & /   \\
\midrule   
\textit{Tacotron 2~\citep{shen2018natural}}    & 3.85 $\pm$ 0.07 & 3.80 $\pm$ 0.08 & 0.115 & 61.78MB & 28.2M \\
\textit{TransformerTTS~\citep{li2018close}}    & 3.87 $\pm$ 0.06 & 3.82 $\pm$ 0.07 & 0.955 & 118.66MB & 24.2M \\
\midrule
\textit{FastSpeech~\citep{ren2019fastspeech}}  & 3.63 $\pm$ 0.08 & 3.72 $\pm$ 0.08 & 0.0198 & 115.2MB & 23.5M \\
\textit{FastSpeech 2~\citep{ren2020fastspeech}}& 3.72 $\pm$ 0.07 & 3.83 $\pm$ 0.06 & 0.0200 & 124.8MB & 27.0M \\
\textit{Glow-TTS~\citep{kim2020glow}}          & 3.61 $\pm$ 0.07 & 3.88 $\pm$ 0.08 & 0.0196 & 116.4MB & 28.6M \\
\textit{BVAE-TTS~\citep{lee2020bidirectional}} & 3.80 $\pm$ 0.06 & 3.72 $\pm$ 0.06 & \textbf{0.0169} & 90.1MB  & 12.0M \\
\midrule 
\textit{PortaSpeech (normal)}                  & \textbf{3.89 $\pm$ 0.06} & \textbf{3.92 $\pm$ 0.06} & 0.0216 & 83.6MB  & 21.8M \\
\textit{PortaSpeech (small)}                   & 3.82 $\pm$ 0.06 & 3.86 $\pm$ 0.06 & 0.0208 & \textbf{39.3MB}  & \textbf{6.7M} \\
\bottomrule
\end{tabular}
\label{tab_main_results}
\end{table}

We compare the quality of generated audio samples, inference latency, model size\footnote{The model parameters do not include the encoder of VAE in BVAE-TTS and PortaSpeech.} and memory footprint\footnote{We profile the peak GPU memory using \textit{MemReporter} in \textit{pytorch\_memlab} (\url{https://github.com/Stonesjtu/pytorch_memlab}) and find the maximum "\textit{active\_bytes}" as the peak memory during inference.} of our PortaSpeech (normal and small model size) with other systems, including 1) \textit{GT}, the ground truth audio; 2) \textit{GT (Mel + HiFi-GAN)}, where we first convert the ground truth audio into mel-spectrograms, and then convert the mel-spectrograms back to audio using HiFi-GAN; 3) \textit{Tacotron 2~\citep{shen2018natural}}; 4) \textit{Transformer TTS~\citep{li2018close}}; 5) \textit{FastSpeech~\citep{ren2019fastspeech}}; 6) \textit{FastSpeech 2~\citep{ren2020fastspeech}}; 7) \textit{Glow-TTS~\citep{kim2020glow}} and 8) \textit{BVAE-TTS~\citep{lee2020bidirectional}}\footnote{We fail in reproducing the performance of BVAE-TTS reported in the original paper, so we use hard text-to-speech alignment in their model and obtain reasonable results.}.  The results are shown in \autoref{tab_main_results}. We have the following observations: 
\begin{itemize}
    \item For \textit{audio quality}, PortaSpeech (normal) outperforms previous TTS models in both audio quality (MOS-Q) and prosody (MOS-P), and only has slight performance degradation when reducing the model size, which shows the superiority of our proposed method.
    \item For \textit{model size} and \textit{memory footprint}, PortaSpeech (small) has the smallest model size and memory footprint. Compared with FastSpeech 2, PortaSpeech (small) achieves 4x model size and 3x memory footprint compression ratios.
    \item For \textit{inference speed}, PortaSpeech (small) speeds up the end-to-end speech generation by 5.5x and 45.9x compared with Tacotron 2 and TransformerTTS and achieves similar RTF with other NAR-TTS models.
\end{itemize}

Besides, we conduct some experiments on the multi-speaker dataset and draw similar conclusions (see Appendix \ref{apdx:multispk}). We also conduct robustness evaluation on both single-speaker and multi-speaker dataset in Appendix \ref{apdx:rob_eval} and find that PortaSpeech achieves comparable robustness performance with state-of-the-art NAR-TTS models.

\subsection{Visualizations}
We then visualize the mel-spectrograms generated by the above systems in \autoref{fig:mel_main_vis}. We can see that PortaSpeech can generate mel-spectrograms with rich details in frequency bins between two adjacent harmonics, unvoiced frames and high-frequency parts, which results in natural sounds. Besides, we visualize the diverse mel-spectrograms generated by PortaSpeech in Appendix \ref{apdx:vis_mels}. In conclusion, our experiments demonstrate that PortaSpeech achieves the goals described in \autoref{sec:intro} (\textit{fast}, \textit{lightweight}, \textit{high-quality}, \textit{expressive} and \textit{diverse}).

\subsection{Ablation Studies}
We conduct ablation studies to demonstrate the effectiveness of designs in PortaSpeech, including the enhanced prior, our post-net and the mixture alignment. We put more analyses on the grouped parameter sharing mechanism in Appendix \ref{apdx:ana_group_psm}. We conduct CMOS evaluation for these ablation studies. The results are shown in \autoref{tab:abl}. 

\paragraph{Enhanced Prior}

\begin{wraptable}{r}{9.0cm}
\centering
\small

\caption{Audio prosody and quality comparisons for ablation study. \textit{MA} denotes mixture alignment in the linguistic encoder; \textit{PN} denotes the flow-based post-net; \textit{EP} denotes the enhanced prior in the variational generator; \textit{Conv} denotes the convolutional post-net used in Tacotron 2~\cite{shen2018natural}.}\label{tab:abl}
\centering
	\begin{tabular}{ l | c | c | c | c}
		\toprule
		\multirow{2}{*}{Setting} & \multicolumn{2}{c|}{normal} & \multicolumn{2}{c}{small} \\\cmidrule{2-5}
		                                  & CMOS-P  & CMOS-Q & CMOS-P  & CMOS-Q  \\
		\midrule
		\textit{PortaSpeech}              & 0.000   & 0.000  & 0.000   & 0.000   \\
		\midrule
		\textit{- EP}                    & -0.194  & -0.014 & -0.212  & -0.098  \\
		\textit{- PN}                    & -0.012  & -0.458 & -0.007  & -0.162  \\
		\textit{- PN + Conv}             & -0.010  & -0.441 & -0.005  & -0.148  \\
		\textit{- MA}                    & -0.241  & -0.127 & -0.312  & -0.157  \\
		\bottomrule
	\end{tabular}
\vspace{-3mm}
\end{wraptable}

To demonstrate the effectiveness of enhanced normalizing flow-based prior, we compare our models with those with simple Gaussian prior as the original VAE. The results are shown in row 2 in \autoref{tab:abl}. We can see that CMOS-P drops when removing the enhanced prior, indicating that the enhanced prior can improve the prosody. Since the prosody is mainly modeled by VAE, compared with simple Gaussian prior, the enhanced prior has weaker assumptions and restrictions on the shape of the VAE prior distribution.

\begin{wraptable}{r}{5.5cm}
    \vspace{-4mm}
    \small
	\caption{Average absolute duration error comparisons in word and sentence level on test set for PortaSpeech (small).}
	\centering
	\begin{tabular}{ l | c | c }
	\toprule
	Settings                  & Word (ms) & Sentences (s)  \\
	\midrule 
	\textit{w/ MA}            & 96.3      & 1.40 \\
	\textit{w/o MA}           & 136.7     & 1.84 \\
	\bottomrule
    \end{tabular}
	\label{tab:dur_acc}
\end{wraptable}

\paragraph{Post-Net} To demonstrate the effectiveness and necessity of flow-based post-net, we compare PortaSpeech with that without the post-net and that with convolutional post-net, which is widely used in previous TTS models, such as Tacotron 2~\cite{shen2018natural}. The results are shown in row 3 and row 4 in \autoref{tab:abl}. From row 3, it can be seen that CMOS-Q drops significantly when removing our post-net, demonstrating that our post-net can improve the audio quality of the generated mel-spectrograms. From row 4, we can see that our flow-based post-net outperforms the commonly used convolutional post-net.

\paragraph{Mixture Alignment}
To demonstrate the effectiveness of mixture alignment, we replace the mixture alignment in the linguistic encoder with the phoneme-level hard alignment proposed in FastSpeech~\cite{ren2019fastspeech}. The results are shown in row 5 in \autoref{tab:abl}. We can see that PortaSpeech with mixture alignment outperforms that with phoneme-level hard alignment in terms of both CMOS-P and CMOS-Q. These results demonstrate that 1) mixture alignment can improve the prosody, which may benefit from more accurate duration extraction and prediction; 2) mixture alignment can also improve the generated voice quality since the soft alignment helps the end-to-end model optimization. Then we calculate the average absolute duration error in word and sentence level on the test set for PortaSpeech (small) with and without mixture alignment. The results are shown in \autoref{tab:dur_acc}. It can be seen that the linguistic encoder with mixture alignment predicts more accurate duration, also demonstrating the effectiveness of the mixture alignment. We visualize the attention alignments generated by our linguistic encoder in Appendix \ref{apdx:ling_enc}, showing that PortaSpeech can create reasonable alignments which is close to the diagonal.

\section{Conclusion}
\label{sec:conslu}
In this paper, we proposed PortaSpeech, a portable and high-quality generative text-to-speech model. PortaSpeech uses a variational generator with an enhanced prior followed by a flow-based post-net with grouped parameter sharing mechanism as the main model architecture. We also proposed a new linguistic encoder with mixture alignment to improve the prosody and reduce the dependence on the hard fine-grained alignment, which combines the hard word-level and soft phoneme-level alignments. Our experimental results show that PortaSpeech outperforms other TTS models in voice quality and prosody and shows only a slight performance degradation when reducing the model size. We also conduct comprehensive ablation studies to verify the effectiveness of each component in PortaSpeech. However, to take advantage of the merits of VAE and normalizing flow, we sacrifice at the cost of more complicated model designs than previous NAR-TTS models: the overall architecture, which cascades linguistic encoder, VAE and post-net, is somewhat complicated. In the future, we will verify the effectiveness of PortaSpeech on multi-speaker and multilingual scenarios. We will also try to tap its potential on other tasks, such as voice conversion and end-to-end text-to-waveform generation.

\section*{Acknowledgments}
This work was supported in part by the National Key R\&D Program of China under Grant No.2020YFC0832505, National Natural Science Foundation of China under Grant No.61836002, No.62072397, Zhejiang Natural Science Foundation under Grant LR19F020006 and Baidu Scholarship Program.

\bibliography{main}

\begin{thebibliography}{10}

\bibitem{alemi2018fixing}
Alexander Alemi, Ben Poole, Ian Fischer, Joshua Dillon, Rif~A Saurous, and
  Kevin Murphy.
\newblock Fixing a broken elbo.
\newblock In {\em International Conference on Machine Learning}, pages
  159--168. PMLR, 2018.

\bibitem{arik2017deep}
Sercan~O Arik, Mike Chrzanowski, Adam Coates, Gregory Diamos, Andrew Gibiansky,
  Yongguo Kang, Xian Li, John Miller, Andrew Ng, Jonathan Raiman, et~al.
\newblock Deep voice: Real-time neural text-to-speech.
\newblock {\em arXiv preprint arXiv:1702.07825}, 2017.

\bibitem{dinh2014nice}
Laurent Dinh, David Krueger, and Yoshua Bengio.
\newblock Nice: Non-linear independent components estimation.
\newblock {\em arXiv preprint arXiv:1410.8516}, 2014.

\bibitem{dinh2016density}
Laurent Dinh, Jascha Sohl-Dickstein, and Samy Bengio.
\newblock Density estimation using real nvp.
\newblock {\em arXiv preprint arXiv:1605.08803}, 2016.

\bibitem{donahue2020end}
Jeff Donahue, Sander Dieleman, Miko{\l}aj Bi{\'n}kowski, Erich Elsen, and Karen
  Simonyan.
\newblock End-to-end adversarial text-to-speech.
\newblock {\em arXiv preprint arXiv:2006.03575}, 2020.

\bibitem{he2019lagging}
Junxian He, Daniel Spokoyny, Graham Neubig, and Taylor Berg-Kirkpatrick.
\newblock Lagging inference networks and posterior collapse in variational
  autoencoders.
\newblock {\em arXiv preprint arXiv:1901.05534}, 2019.

\bibitem{ljspeech17}
Keith Ito.
\newblock The lj speech dataset.
\newblock \url{https://keithito.com/LJ-Speech-Dataset/}, 2017.

\bibitem{kim2020glow}
Jaehyeon Kim, Sungwon Kim, Jungil Kong, and Sungroh Yoon.
\newblock Glow-tts: A generative flow for text-to-speech via monotonic
  alignment search.
\newblock {\em arXiv preprint arXiv:2005.11129}, 2020.

\bibitem{kingma2018glow}
Durk~P Kingma and Prafulla Dhariwal.
\newblock Glow: Generative flow with invertible 1x1 convolutions.
\newblock In {\em Advances in Neural Information Processing Systems}, pages
  10215--10224, 2018.

\bibitem{kingma2016improved}
Durk~P Kingma, Tim Salimans, Rafal Jozefowicz, Xi~Chen, Ilya Sutskever, and Max
  Welling.
\newblock Improved variational inference with inverse autoregressive flow.
\newblock {\em Advances in neural information processing systems},
  29:4743--4751, 2016.

\bibitem{kong2020hifi}
Jungil Kong, Jaehyeon Kim, and Jaekyoung Bae.
\newblock Hifi-gan: Generative adversarial networks for efficient and high
  fidelity speech synthesis.
\newblock {\em Advances in Neural Information Processing Systems}, 33, 2020.

\bibitem{lancucki2020fastpitch}
Adrian {\L}a{\'n}cucki.
\newblock Fastpitch: Parallel text-to-speech with pitch prediction.
\newblock {\em arXiv preprint arXiv:2006.06873}, 2020.

\bibitem{lee2020nanoflow}
Sang-gil Lee, Sungwon Kim, and Sungroh Yoon.
\newblock Nanoflow: Scalable normalizing flows with sublinear parameter
  complexity.
\newblock {\em arXiv preprint arXiv:2006.06280}, 2020.

\bibitem{lee2020bidirectional}
Yoonhyung Lee, Joongbo Shin, and Kyomin Jung.
\newblock Bidirectional variational inference for non-autoregressive
  text-to-speech.
\newblock In {\em International Conference on Learning Representations}, 2020.

\bibitem{li2018close}
Naihan Li, Shujie Liu, Yanqing Liu, Sheng Zhao, and Ming Liu.
\newblock Neural speech synthesis with transformer network.
\newblock In {\em Proceedings of the AAAI Conference on Artificial
  Intelligence}, volume~33, pages 6706--6713, 2019.

\bibitem{lim2020jdi}
Dan Lim, Won Jang, Hyeyeong Park, Bongwan Kim, Jesam Yoon, et~al.
\newblock Jdi-t: Jointly trained duration informed transformer for
  text-to-speech without explicit alignment.
\newblock {\em arXiv preprint arXiv:2005.07799}, 2020.

\bibitem{liu2021diffsinger}
Jinglin Liu, Chengxi Li, Yi~Ren, Feiyang Chen, Peng Liu, and Zhou Zhao.
\newblock Diffsinger: Singing voice synthesis via shallow diffusion mechanism.
\newblock {\em arXiv preprint arXiv:2105.02446}, 2, 2021.

\bibitem{luo2021lightspeech}
Renqian Luo, Xu~Tan, Rui Wang, Tao Qin, Jinzhu Li, Sheng Zhao, Enhong Chen, and
  Tie-Yan Liu.
\newblock Lightspeech: Lightweight and fast text to speech with neural
  architecture search.
\newblock {\em arXiv preprint arXiv:2102.04040}, 2021.

\bibitem{mahajan2020latent}
Shweta Mahajan, Iryna Gurevych, and Stefan Roth.
\newblock Latent normalizing flows for many-to-many cross-domain mappings.
\newblock {\em arXiv preprint arXiv:2002.06661}, 2020.

\bibitem{miao2020flow}
Chenfeng Miao, Shuang Liang, Minchuan Chen, Jun Ma, Shaojun Wang, and Jing
  Xiao.
\newblock Flow-tts: A non-autoregressive network for text to speech based on
  flow.
\newblock In {\em ICASSP 2020-2020 IEEE International Conference on Acoustics,
  Speech and Signal Processing (ICASSP)}, pages 7209--7213. IEEE, 2020.

\bibitem{ming2016deep}
Huaiping Ming, Dongyan Huang, Lei Xie, Jie Wu, Minghui Dong, and Haizhou Li.
\newblock Deep bidirectional lstm modeling of timbre and prosody for emotional
  voice conversion.
\newblock 2016.

\bibitem{peng2020non}
Kainan Peng, Wei Ping, Zhao Song, and Kexin Zhao.
\newblock Non-autoregressive neural text-to-speech.
\newblock ICML, 2020.

\bibitem{ping2018deep}
Wei Ping, Kainan Peng, Andrew Gibiansky, Sercan~O. Arik, Ajay Kannan, Sharan
  Narang, Jonathan Raiman, and John Miller.
\newblock Deep voice 3: 2000-speaker neural text-to-speech.
\newblock In {\em International Conference on Learning Representations}, 2018.

\bibitem{prenger2019waveglow}
Ryan Prenger, Rafael Valle, and Bryan Catanzaro.
\newblock Waveglow: A flow-based generative network for speech synthesis.
\newblock In {\em ICASSP 2019-2019 IEEE International Conference on Acoustics,
  Speech and Signal Processing (ICASSP)}, pages 3617--3621. IEEE, 2019.

\bibitem{ren2020fastspeech}
Yi~Ren, Chenxu Hu, Tao Qin, Sheng Zhao, Zhou Zhao, and Tie-Yan Liu.
\newblock Fastspeech 2: Fast and high-quality end-to-end text-to-speech.
\newblock {\em arXiv preprint arXiv:2006.04558}, 2020.

\bibitem{ren2019fastspeech}
Yi~Ren, Yangjun Ruan, Xu~Tan, Tao Qin, Sheng Zhao, Zhou Zhao, and Tie-Yan Liu.
\newblock Fastspeech: Fast, robust and controllable text to speech.
\newblock In {\em Advances in Neural Information Processing Systems}, pages
  3165--3174, 2019.

\bibitem{rezende2015variational}
Danilo~Jimenez Rezende and Shakir Mohamed.
\newblock Variational inference with normalizing flows.
\newblock {\em arXiv preprint arXiv:1505.05770}, 2015.

\bibitem{setiawan2020variational}
Hendra Setiawan, Matthias Sperber, Udhay Nallasamy, and Matthias Paulik.
\newblock Variational neural machine translation with normalizing flows.
\newblock {\em arXiv preprint arXiv:2005.13978}, 2020.

\bibitem{shaw2018self}
Peter Shaw, Jakob Uszkoreit, and Ashish Vaswani.
\newblock Self-attention with relative position representations.
\newblock {\em arXiv preprint arXiv:1803.02155}, 2018.

\bibitem{shen2018natural}
Jonathan Shen, Ruoming Pang, Ron~J Weiss, Mike Schuster, Navdeep Jaitly,
  Zongheng Yang, Zhifeng Chen, Yu~Zhang, Yuxuan Wang, Rj~Skerrv-Ryan, et~al.
\newblock Natural tts synthesis by conditioning wavenet on mel spectrogram
  predictions.
\newblock In {\em 2018 IEEE International Conference on Acoustics, Speech and
  Signal Processing (ICASSP)}, pages 4779--4783. IEEE, 2018.

\bibitem{sun2019token}
Hao Sun, Xu~Tan, Jun-Wei Gan, Hongzhi Liu, Sheng Zhao, Tao Qin, and Tie-Yan
  Liu.
\newblock Token-level ensemble distillation for grapheme-to-phoneme conversion.
\newblock In {\em INTERSPEECH}, 2019.

\bibitem{tomczak2018vae}
Jakub Tomczak and Max Welling.
\newblock Vae with a vampprior.
\newblock In {\em International Conference on Artificial Intelligence and
  Statistics}, pages 1214--1223. PMLR, 2018.

\bibitem{vainer2020speedyspeech}
Jan Vainer and Ond{\v{r}}ej Du{\v{s}}ek.
\newblock Speedyspeech: Efficient neural speech synthesis.
\newblock {\em arXiv preprint arXiv:2008.03802}, 2020.

\bibitem{van2016wavenet}
A{\"a}ron Van Den~Oord, Sander Dieleman, Heiga Zen, Karen Simonyan, Oriol
  Vinyals, Alex Graves, Nal Kalchbrenner, Andrew~W Senior, and Koray
  Kavukcuoglu.
\newblock Wavenet: A generative model for raw audio.
\newblock {\em SSW}, 125, 2016.

\bibitem{vaswani2017attention}
Ashish Vaswani, Noam Shazeer, Niki Parmar, Jakob Uszkoreit, Llion Jones,
  Aidan~N Gomez, {\L}ukasz Kaiser, and Illia Polosukhin.
\newblock Attention is all you need.
\newblock In {\em Advances in Neural Information Processing Systems}, pages
  5998--6008, 2017.

\bibitem{wang2017tacotron}
Yuxuan Wang, RJ~Skerry-Ryan, Daisy Stanton, Yonghui Wu, Ron~J Weiss, Navdeep
  Jaitly, Zongheng Yang, Ying Xiao, Zhifeng Chen, Samy Bengio, et~al.
\newblock Tacotron: Towards end-to-end speech synthesis.
\newblock {\em arXiv preprint arXiv:1703.10135}, 2017.

\bibitem{yamamoto2020parallel}
Ryuichi Yamamoto, Eunwoo Song, and Jae-Min Kim.
\newblock Parallel wavegan: A fast waveform generation model based on
  generative adversarial networks with multi-resolution spectrogram.
\newblock In {\em ICASSP 2020-2020 IEEE International Conference on Acoustics,
  Speech and Signal Processing (ICASSP)}, pages 6199--6203. IEEE, 2020.

\end{thebibliography}
\bibliographystyle{plain}

\newpage
\appendix 
\appendixpage
\section{Details of Models}
\label{apdx:model_details}
In this section, we describe details in the linguistic encoder, variational generator, post-net and the models we used in \autoref{sec:ana_vae_flow}.
\subsection{Linguistic Encoder}
\label{apdx:ling_enc}
\begin{figure}[!h]
	\centering
	\begin{subfigure}[h]{0.24\textwidth}
		\centering
		\includegraphics[width=\textwidth,trim={0cm 0.05cm 9cm 0cm}, clip=true]{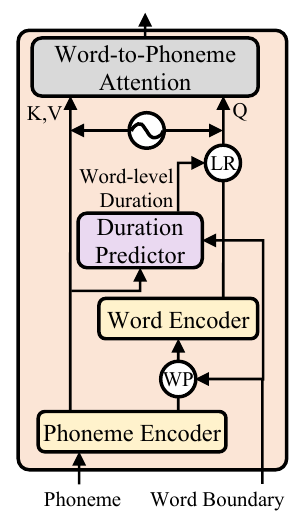}
		\caption{Linguistic Encoder}
		\label{fig:lin_encoder}
	\end{subfigure}
	\begin{subfigure}[h]{0.25\textwidth}
		\centering
		\includegraphics[width=\textwidth,trim={3.2cm 0.05cm 5.8cm 0.2cm}, clip=true]{figs/arch_lin_encoder.pdf}
		\caption{Phoneme/Word Encoder}
		\label{fig:ph_encoder}
	\end{subfigure}
	\begin{subfigure}[h]{0.22\textwidth}
		\centering
		\includegraphics[width=\textwidth,trim={6.3cm 0.0cm 3.0cm 0.1cm}, clip=true]{figs/arch_lin_encoder.pdf}
		\caption{Duration Predictor}
		\label{fig:dur_pred}
	\end{subfigure}
	\begin{subfigure}[h]{0.23\textwidth}
		\centering
		\includegraphics[width=\textwidth,trim={9.0cm 0.0cm 0.1cm 0.0cm}, clip=true]{figs/arch_lin_encoder.pdf}
		\caption{Word-Level Pooling}
		\label{fig:wl_pool}
	\end{subfigure}
	\caption{The detailed architecture of linguistic encoder.}
	\label{fig:arch_lin_encoder}
\end{figure}

As shown in \autoref{fig:arch_lin_encoder}, our linguistic encoder consists of a phoneme encoder, a word encoder, a duration predictor and a word-to-phoneme attention module. \textbf{The phoneme encoder and the word encoder} are both stacks of feed-forward Transformer layers with relative position encoding~\cite{shaw2018self}, as shown in \autoref{fig:ph_encoder}. \textbf{The duration predictor}, as shown in \autoref{fig:dur_pred}, consists of two 1D-convolutional layers, each of which is followed by ReLU activation and layer normalization, and a linear layer to project the hidden states in each timestep to a scalar, which is the predicted phoneme duration. \textbf{The word-level pooling} averages the phoneme hidden states inside each word according to the word boundary, as shown in \autoref{fig:wl_pool}. \textbf{The word-to-phoneme attention module} is a multi-head attention~\cite{vaswani2017attention} with 2 heads and we apply a word-to-phoneme mapping mask to the attention weight to force each query (Q) to only attend to the phonemes belongs to the word corresponding to this query. We also add a well-designed \textbf{positional encoding} to the inputs of word-to-phoneme attention module: for K and V, the positional encoding is: $\frac{i}{L_w}E_{kv}$, where $i$ is the position of the corresponding phoneme in the word $w$; $L_w$ is the number of phonemes in word $w$; $E_{kv}$ is a learnable embedding; and $i \in \{0, 1, ..., L_w-1\}$. For Q, the positional encoding becomes: $\frac{j}{T_w}E_{q}$, where $j$ is the position of the corresponding frame in the word $w$; $T_w$ is the number of frames in word $w$; $E_{q}$ is another learnable embedding; and $j \in \{0, 1, ..., T_w-1\}$.

\subsection{Variational Generator}
\label{apdx:vg}
\begin{figure}[!h]
	\centering
	\begin{subfigure}[h]{0.26\textwidth}
		\centering
		\includegraphics[width=\textwidth,trim={0cm 0cm 9.2cm 0cm}, clip=true]{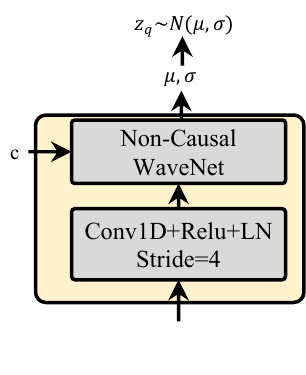}
		\caption{Encoder}
		\label{fig:vg_encoder}
	\end{subfigure}
	\begin{subfigure}[h]{0.26\textwidth}
		\centering
		\includegraphics[width=\textwidth,trim={3.2cm -0.1cm 5.85cm -0.1cm}, clip=true]{figs/arch_vg.pdf}
		\caption{Decoder}
		\label{fig:vg_decoder}
	\end{subfigure}
	\begin{subfigure}[h]{0.24\textwidth}
		\centering
		\includegraphics[width=\textwidth,trim={6.5cm -0.1cm 2.8cm -0.1cm}, clip=true]{figs/arch_vg.pdf}
		\caption{VP Flow}
		\label{fig:vg_prior}
	\end{subfigure}
	\begin{subfigure}[h]{0.22\textwidth}
		\centering
		\includegraphics[width=\textwidth,trim={9.6cm 0cm 0.1cm 0.0cm}, clip=true]{figs/arch_vg.pdf}
		\caption{Coupling Layer}
		\label{fig:vg_rsl}
	\end{subfigure}
	\caption{The detailed architecture of variational generator.}
	\label{fig:arch_vg}
\end{figure}

As shown in \autoref{fig:arch_vg}, our variational generator consists of an encoder, a decoder and a volume-preserving (VP) flow-based prior model. \textbf{The encoder}, as shown in \autoref{fig:vg_encoder}, is composed of a 1D-convolution with stride 4 followed by ReLU activation and layer normalization, and a non-causal WaveNet. \textbf{The decoder}, as shown in \autoref{fig:vg_decoder}, consists of a non-causal WaveNet and a 1D transposed convolution with stride 4, also followed by ReLU and layer normalization. \textbf{The prior model}, as shown in \autoref{fig:vg_prior}, is a volume-preserving normalizing flow, which is composed of a residual coupling layer (\autoref{fig:vg_rsl}) and a channel-wise flip operation.

\subsection{Post-Net}
\label{apdx:pn}
We use non-causal WaveNet as the main architecture of NN in the affine coupling layer. We introduce the number of shared groups $N_g$, for example, when $N_g=2$, NNs in flow steps ($\bbf_1$, $\bbf_{2}$, ..., $\bbf_{K/2}$) and ($\bbf_{K/2+1}$, $\bbf_{K/2+2}$, ..., $\bbf_{K}$) share the parameters separately. In inference, we can sample $z$ from $N(0, T^2)$, where $T$ is the temperature and use $T=0.8$ by default.

\subsection{Models Used in Section 4.2}
\label{apdx:model_ana}
\begin{figure}[!h]
	\centering
	\begin{subfigure}[h]{0.36\textwidth}
		\centering
		\includegraphics[width=0.85\textwidth,trim={0cm 0.4cm 5.8cm 0cm}, clip=true]{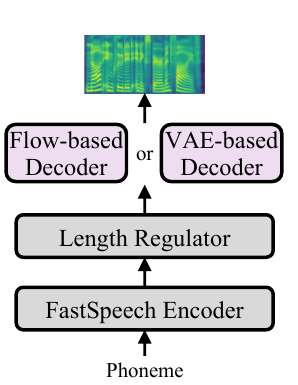}
		\caption{TTS with VAE/flow-based Decoder}
		\label{fig:arch_vae_flow}
	\end{subfigure}
	\begin{subfigure}[h]{0.28\textwidth}
		\centering
		\includegraphics[width=\textwidth,trim={3cm -0.4cm 2.5cm 0cm}, clip=true]{figs/arch_analyses_vae_glow.pdf}
		\caption{Flow-based Decoder}
		\label{fig:arch_flow_dec}
	\end{subfigure}
	\begin{subfigure}[h]{0.27\textwidth}
		\centering
		\includegraphics[width=0.9\textwidth,trim={6.5cm 0.0cm 0cm 0.5cm}, clip=true]{figs/arch_analyses_vae_glow.pdf}
		\caption{VAE-based Decoder}
		\label{fig:arch_vae_dec}
	\end{subfigure}
	\caption{The detailed architecture of NAR-TTS models with VAE and flow-based decoders.}
	\label{fig:vae_flow}
\end{figure}

We use FastSpeech~\cite{ren2019fastspeech} as the backbone for preliminary analyses in \autoref{sec:ana_vae_flow}. We replace the decoder of FastSpeech with flow-based decoder and VAE-based decoder to explore the characteristics of them. The flow-based decoder is mainly adopted from Glow~\cite{kingma2018glow} and WaveGlow~\cite{prenger2019waveglow}, which uses the expanded encoder outputs as the condition, as shown in \autoref{fig:arch_vae_flow}. The VAE-based decoder is similar to the variational generator in our proposed PortaSpeech, except that it does not use the flow-based prior. The model hyperparameters of different model configurations are listed in \autoref{tab:hyperparameters_ana}.

\begin{table}[h]
\vspace{-2mm}
\caption{Hyperparameters of VAE and flow-based TTS models.}
\vspace{2mm}
\label{tab:hyperparameters_ana}
\small
\centering
\begin{tabular}{l|l|c|c|c|c|c|c}
\toprule
& \multirow{2}{*}{Hyperparameter} & \multicolumn{3}{c|}{Flow-based} & \multicolumn{3}{c}{VAE-based} \\\cmidrule{3-8}
                           & & big & middle & small & big & middle & small \\ \midrule

\multirow{5}{*}{Encoder} 
&Phoneme Embedding           & 256 & 192    & 128   & 256 & 192    & 128   \\
&Layers                      & 4   & 4      & 3     & 4   & 4      & 3     \\
&Hidden Size                 & 256 & 192    & 128   & 256 & 192    & 128   \\
&Conv1D Kernel               & 9   & 5      & 3     & 9   & 5      & 3     \\
&Conv1D Filter Size          & 1024& 768    & 512   & 1024& 768    & 512   \\ \midrule
\multirow{5}{*}{VAE Decoder}
&VAE Encoder Layers          &\multicolumn{3}{c|}{/}&\multicolumn{3}{c}{8} \\ 
&VAE Conv1D Kernel           &\multicolumn{3}{c|}{/}&\multicolumn{3}{c}{5} \\ 
&Latent Size                 &\multicolumn{3}{c|}{/}&\multicolumn{3}{c}{16}\\ 
&WaveNet Channel Size        &\multicolumn{3}{c|}{/}& 300 & 128    & 128    \\ 
&VAE Decoder Layers          &\multicolumn{3}{c|}{/}& 16  & 12     & 12     \\ 
\midrule
\multirow{4}{*}{Flow Decoder}
&WaveNet Layers              &\multicolumn{3}{c|}{4}& \multicolumn{3}{c}{/}    \\
&WaveNet Kernel              &\multicolumn{3}{c|}{5}& \multicolumn{3}{c}{/}    \\
&WaveNet Channel Size        & 128 & 112    & 112   & \multicolumn{3}{c}{/}   \\ 
&Flow Steps                  & 22  & 6      & 4     & \multicolumn{3}{c}{/}   \\ 
\midrule
\multicolumn{2}{c|}{Total Number of Parameters} 
                             &41.2M&10.2M  &4.5M   &43.2M &9.3M    &4.4M       \\
\bottomrule
\end{tabular}
\end{table}

\section{Detailed Experimental Settings}
\label{apdx:exp_settings}
In this section, we describe more model configurations and details in subjective evaluation.

\subsection{Model Configurations}
We list the model hyper-parameters of PortaSpeech (normal) and PortaSpeech (small) in \autoref{tab:hyperparameters_ps} and total number of parameters of each module in \autoref{tab:num_params_module}. 

\begin{table}[h]
\vspace{-2mm}
\caption{Hyperparameters of PortaSpeech (normal) and PortaSpeech (small) models.}
\vspace{2mm}
\label{tab:hyperparameters_ps}
\small
\centering
\begin{tabular}{l|l|c|c}
\toprule
\multicolumn{2}{c|}{Hyperparameter} & PortaSpeech (normal) & PortaSpeech (small) \\ \midrule
\multirow{5}{*}{Linguistic Encoder} 
&Phoneme Embedding            & 192  & 128      \\
&Word/Phoneme Encoder Layers  & 4    & 3        \\
&Hidden Size                  & 192  & 128      \\
&Conv1D Kernel                & 5    & 3        \\
&Conv1D Filter Size           & 768  & 512      \\ 
\midrule
\multirow{9}{*}{Varational Generator}
&Encoder Layers               & 8    & 8  \\ 
&Encoder Kernel               & 5    & 3  \\ 
&Decoder Layers               & 4    & 3  \\ 
&Encoder/Decoder Kernel       & 5    & 3  \\ 
&Encoder/Decoder Channel Size & 192  & 128 \\ 
&Latent Size                  & 16   & 16  \\ 
&VP-Flow Steps                & 4    & 3  \\
&VP-Flow Layers               & 4    & 4  \\
&VP-Flow Channel Size         & 64   & 32 \\
&VP-Flow Conv1D Kernel        & 3    & 3  \\
\midrule 
\multirow{4}{*}{Post-Net}
&WaveNet Layers               & 3    & 3  \\
&WaveNet Kernel               & 3    & 3  \\
&WaveNet Channel Size         & 192  & 128\\ 
&Flow Steps                   & 12   & 8  \\ 
&Shared Groups                & 3    & 2  \\ 
\midrule
\multicolumn{2}{c|}{Total Number of Parameters}  & 21.8M & 6.7M     \\
\bottomrule
\end{tabular}
\end{table}

\begin{table}[h]
\vspace{-2mm}
\caption{Total number of parameters of each module in PortaSpeech (normal) and PortaSpeech (small).}
\vspace{2mm}
\label{tab:num_params_module}
\small
\centering
\begin{tabular}{l|c|c}
\toprule
Modules             &  PortaSpeech (normal) & PortaSpeech (small)  \\ \midrule
Linguistic Encoder  & 7.2M            & 2.0M           \\
Duration predictor  & 0.3M            & 0.2M           \\
Post-Net            & 10.8M           & 3.6M           \\
Decoder in VG       & 2.5M            & 0.6M           \\
VP-Flow in VG       & 1.0M            & 0.3M           \\ \midrule
Total               & 21.8M           & 6.7M           \\
\bottomrule
\end{tabular}
\end{table}

\subsection{Details in Subjective Evaluation}
\label{apdx:suj_eval_details}

\begin{figure}[!h]
	\centering
	\begin{subfigure}[h]{\textwidth}
		\centering
		\includegraphics[width=\textwidth,trim={0cm 0cm 0cm 0cm}, clip=true]{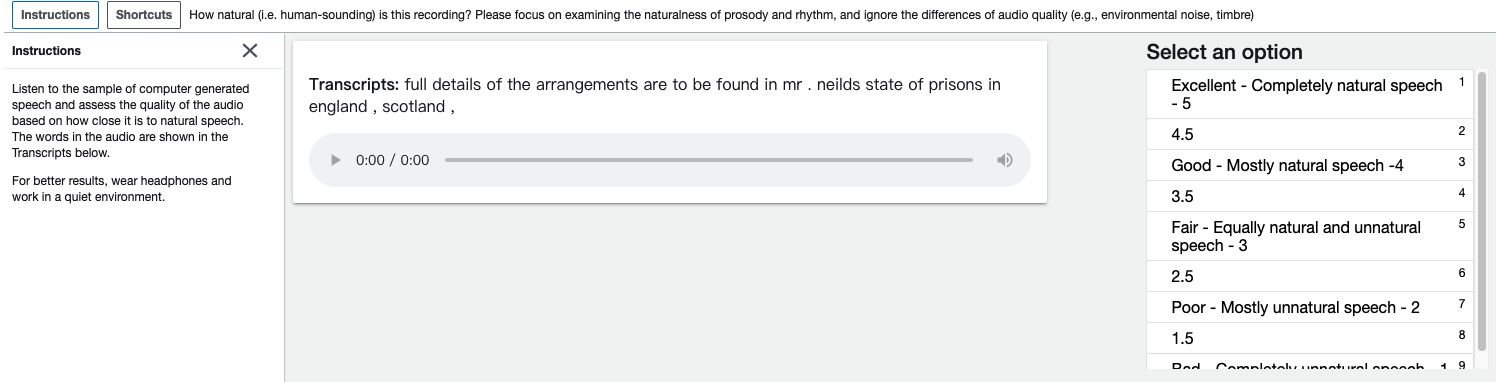}
		\caption{Screenshot of MOS-P testing.}
		\label{fig:screenshot_mos_p}
	\end{subfigure}
	\begin{subfigure}[h]{\textwidth}
		\centering
		\includegraphics[width=\textwidth,trim={0cm 0cm 0cm 0cm}, clip=true]{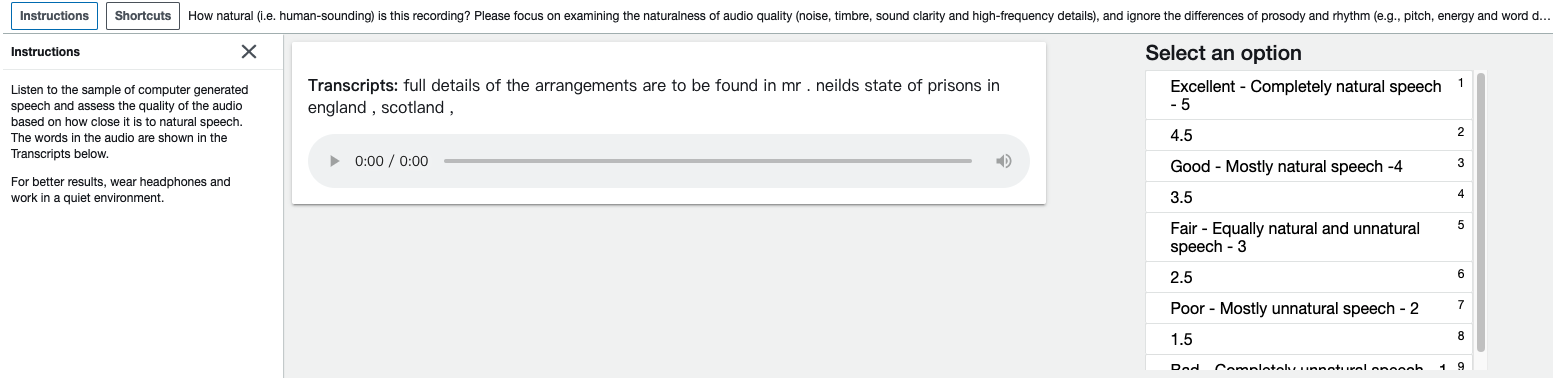}
		\caption{Screenshot of MOS-Q testing.}
		\label{fig:screenshot_mos_Q}
	\end{subfigure}
	\begin{subfigure}[h]{\textwidth}
		\centering
		\includegraphics[width=\textwidth,trim={0cm 0cm 0cm 0cm}, clip=true]{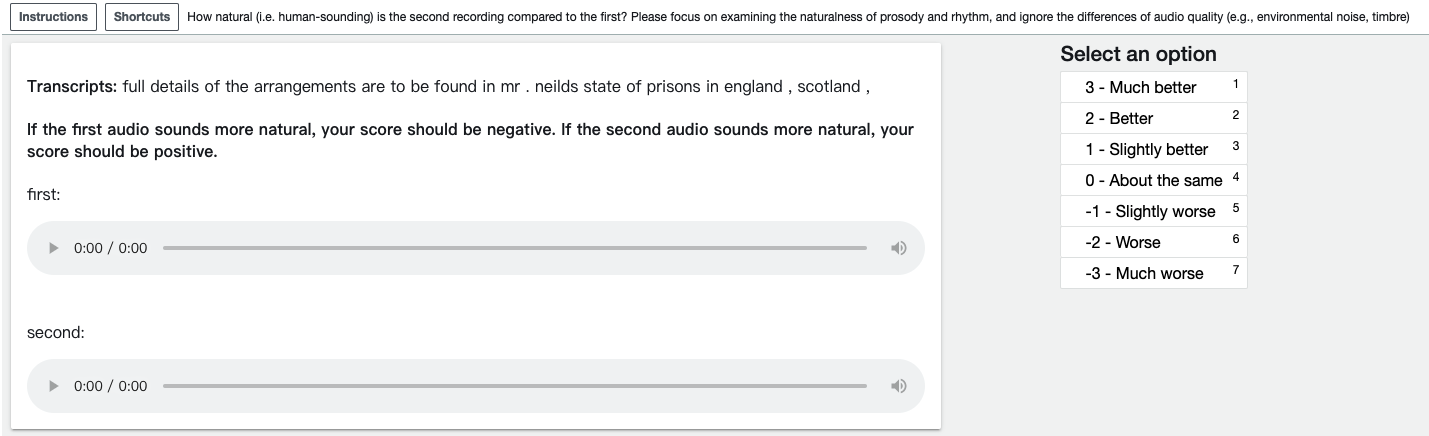}
		\caption{Screenshot of CMOS-P testing.}
		\label{fig:screenshot_cmos_p}
	\end{subfigure}
	\begin{subfigure}[h]{\textwidth}
		\centering
		\includegraphics[width=\textwidth,trim={0cm 0cm 0cm 0cm}, clip=true]{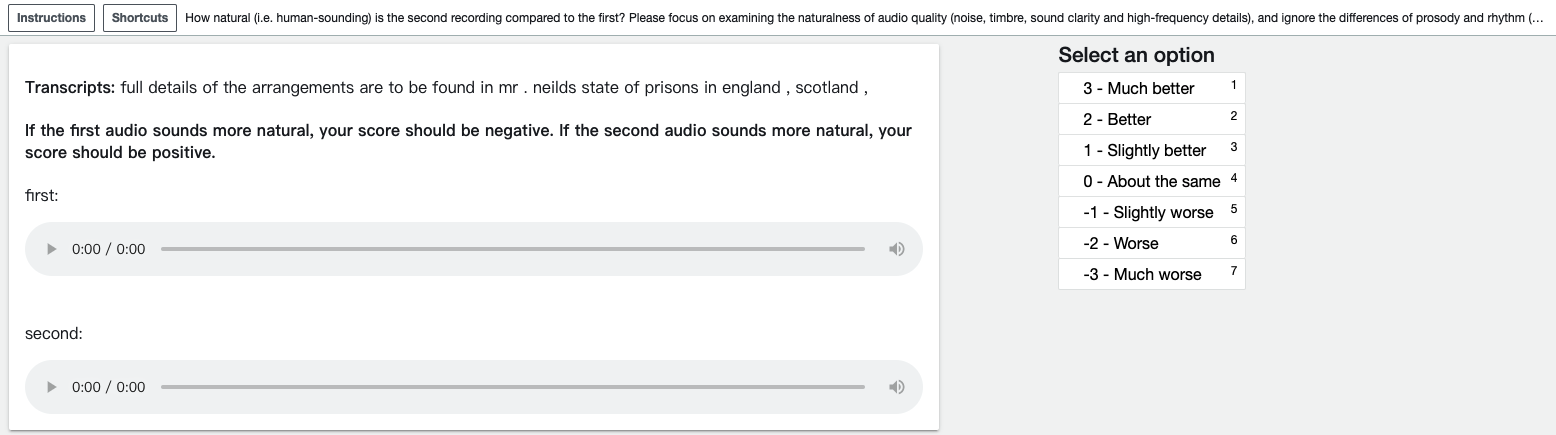}
		\caption{Screenshot of CMOS-Q testing.}
		\label{fig:screenshot_cmos_Q}
	\end{subfigure}
	\caption{Screenshots of subjective evaluations.}
	\label{fig:screenshot_eval}
\end{figure}

For MOS, each tester is asked to evaluate the subjective naturalness of a sentence on a 1-5 Likert scale. For CMOS, listeners are asked to compare pairs of audio generated by systems A and B and indicate which of the two audio they prefer and choose one of the following scores: 0 indicating no difference, 1 indicating small difference, 2 indicating a large difference and 3 indicating a very large difference. For audio quality evaluation (MOS-Q and CMOS-Q), we tell listeners to "\textit{focus on examining the naturalness of prosody and rhythm, and ignore the differences of audio quality (e.g., environmental noise, timbre)}". For prosody evaluations (MOS-P and CMOS-P), we tell listeners to "\textit{focus on examining the naturalness of prosody and rhythm, and ignore the differences of audio quality (e.g., environmental noise, timbre)}". The screenshots of instructions for testers are shown in \autoref{fig:screenshot_eval}. We paid \$8 to participants hourly and totally spent about \$750 on participant compensation.

\section{Results on Multi-Speaker Dataset}
\label{apdx:multispk}
We conduct the MOS evaluation on the multi-speaker dataset: LibriTTS. The results are shown in \autoref{tab:res_libritts} (we use a pre-trained Parallel WaveGAN~\cite{yamamoto2020parallel} for LibriTTS as the vocoder). We can draw similar conclusions as that on LJSpeech that PortaSpeech can achieve good prosody and audio quality in terms of MOS-P and MOS-Q, even in more complicated (multi-speaker) scenarios.

\begin{table}[!h]
\centering
\vspace{-2mm}
\caption{The audio performance (MOS-Q and MOS-P) comparisons on LibriTTS dataset.}
\vspace{2mm}
\begin{tabular}{ l | c | c  }
\toprule
\textbf{Method}        & MOS-P     & MOS-Q    \\ 
\midrule
GT                     & 4.24$\pm$0.08          & 4.36$\pm$0.09          \\
GT   (vocoder)         & 4.21$\pm$0.09          & 4.01$\pm$0.10          \\
Tacotron   2           & 3.81$\pm$0.10          & 3.71$\pm$0.11          \\
TransformerTTS         & 3.79$\pm$0.09          & 3.72$\pm$0.12          \\
FastSpeech             & 3.59$\pm$0.11          & 3.61$\pm$0.14          \\
FastSpeech   2         & 3.64$\pm$0.11          & 3.70$\pm$0.11          \\
Glow-TTS               & 3.76$\pm$0.15          & 3.78$\pm$0.10          \\
\midrule
PortaSpeech   (normal) & \textbf{3.84$\pm$0.13} & \textbf{3.83$\pm$0.13} \\
PortaSpeech   (small)  & 3.80$\pm$0.12          & 3.81$\pm$0.11          \\ 
\bottomrule
\end{tabular}
\label{tab:res_libritts}
\end{table}

\section{Robustness Evaluation}
\label{apdx:rob_eval}
We conduct the robustness evaluation on LJSpeech and LibriTTS datasets. We select 50 sentences that are particularly hard for TTS systems following FastSpeech~\cite{ren2019fastspeech}. The results are shown in Tables \ref{tab:rob_ljspeech} and \ref{tab:rob_lib}. We can see that PortaSpeech achieves comparable robustness performance with state-of-the-art non-autoregressive TTS models.

\begin{table}[!h]
\centering
\vspace{-2mm}
\caption{The robustness evaluation on LJSpeech dataset.}
\vspace{2mm}
\begin{tabular}{ l | c | c | c }
\toprule
Method                 & Repeats   & Skips         & Error Sentences \\ \midrule
Tacotron 2             & 4                  & 5                  & 7 \\
TransformerTTS         & 7                  & 7                  & 9 \\
FastSpeech             & 0                  & 1                  & 1 \\
FastSpeech 2           & 0                  & 1                  & 1 \\
Glow-TTS               & 0                  & 2                  & 2 \\ 
\midrule
PortaSpeech (normal)   & 1                  & 0                  & 1 \\
PortaSpeech (small)    & 1                  & 1                  & 1 \\ 
\bottomrule
\end{tabular}
\label{tab:rob_ljspeech}
\end{table}

\begin{table}[!h]
\centering
\vspace{-2mm}
\caption{The robustness evaluation on LibriTTS dataset.}
\vspace{2mm}
\begin{tabular}{ l | c | c | c }
\toprule
Method                 & Repeats   & Skips           & Error Sentences \\ \midrule
Tacotron 2             & 6                  & 7                  & 12  \\
TransformerTTS         & 10                 & 12                 & 15  \\
FastSpeech             & 2                  & 1                  & 2   \\
FastSpeech 2           & 2                  & 1                  & 2   \\
Glow-TTS               & 5                  & 4                  & 8   \\ 
\midrule
PortaSpeech (normal)   & 1                  & 2                  & 2   \\
PortaSpeech (small)    & 2                  & 2                  & 2   \\ 
\bottomrule
\end{tabular}
\label{tab:rob_lib}
\end{table}

\section{Visualization of Attention Weights}
\label{apdx:vis_attn}
We put some word-to-phoneme attention visualizations in \autoref{fig:vis_attn}. We can see that PortaSpeech can create reasonable phoneme-to-spectrogram alignments which are close to the diagonal, which helps the end-to-end training.

\newcommand{\plotattn}[2]{
    \begin{subfigure}{0.235\textwidth}
	\centering
	\includegraphics[width=\textwidth,trim={2mm 2mm 2mm 2mm}, clip=true]{#1}
	\caption{\textit{#2}}
    \end{subfigure}
}
\begin{figure}[!h]
    \centering
    \plotattn{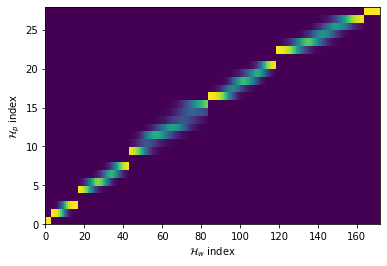}{LJ001-0002}
    \plotattn{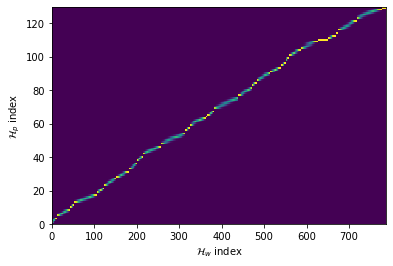}{LJ001-0003}
    \plotattn{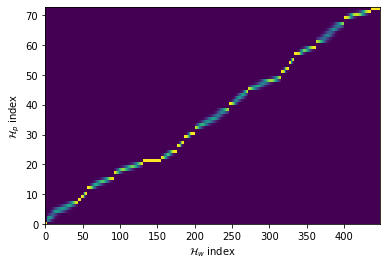}{LJ001-0004}
    \plotattn{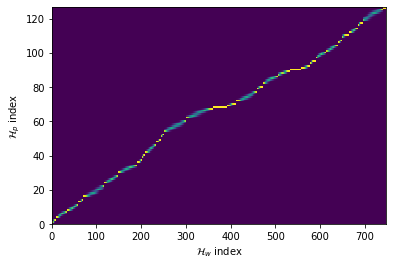}{LJ001-0005}
    \caption{Visualizations of the attention weights.}
    \label{fig:vis_attn}
\end{figure}

\section{More Visualizations of Mel-Spectrograms}
\label{apdx:vis_mels}
We put more visualizations of mel-spectrograms with different sampling temperatures of post-net and different random seeds on PortaSpeech (normal) in \autoref{fig:vis_temp} and \autoref{fig:vis_seed}. We have several observations: 1) From \autoref{fig:vis_temp}, we can see that when $T=0.8$, our model can generate natural sound perceptually with reasonable details in mel-spectrograms. 2) From \autoref{fig:vis_seed}, we can see that with different random seeds, PortaSpeech can generate diverse results, which have different prosody and mel-spectrogram details.

\newcommand{\plotmelLL}[2]{
    \begin{subfigure}{0.320\textwidth}
	\centering
	\includegraphics[width=\textwidth,trim={2mm 2mm 2mm 2mm}, clip=true]{#1}
	\caption{\textit{#2}}
    \end{subfigure}
}
\begin{figure}[!h]
    \centering
    \plotmelLL{figs/mel_plots_main/gt.png}{$GT$}
    \plotmelLL{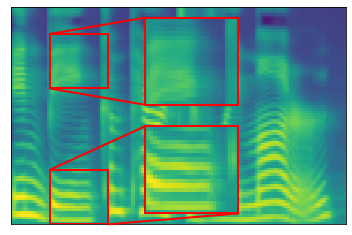}{$T=0.2$}
    \plotmelLL{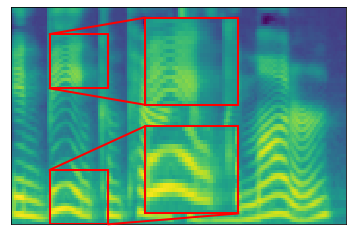}{$T=0.4$}\\
    \plotmelLL{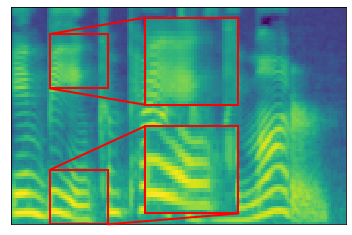}{$T=0.6$}
    \plotmelLL{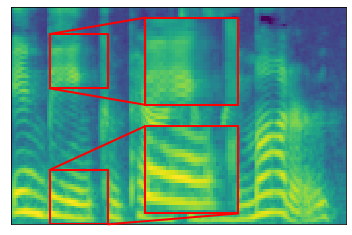}{$T=0.8$}
    \plotmelLL{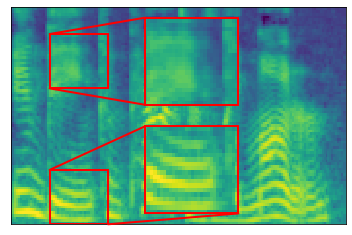}{$T=1.0$}
    \caption{Visualizations of the ground-truth and generated mel-spectrograms generated with different sampling temperature $T$ of post-net.}
    \label{fig:vis_temp}
\end{figure}

\begin{figure}[!h]
    \centering
    \plotmelLL{figs/mel_plots_main/gt.png}{$GT$}
    \plotmelLL{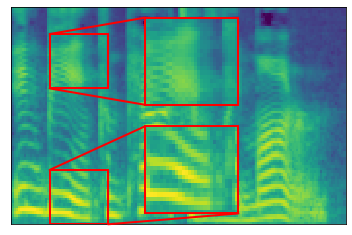}{$S=1237$}
    \plotmelLL{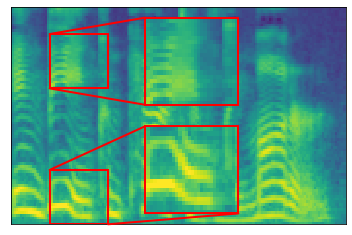}{$S=1239$}\\
    \plotmelLL{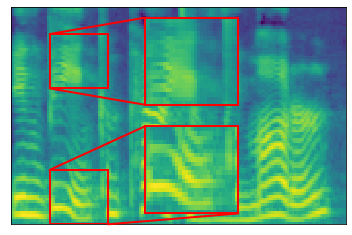}{$S=3237$}
    \plotmelLL{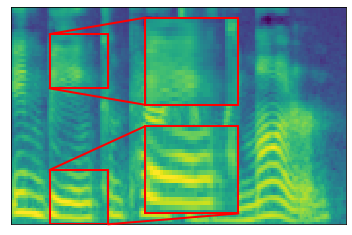}{$S=4237$}
    \plotmelLL{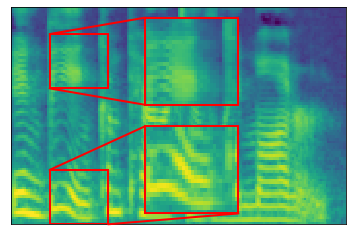}{$S=9237$}
    \caption{Visualizations of the ground-truth and generated mel-spectrograms generated with different random seeds $S$.}
    \label{fig:vis_seed}
\end{figure}

\section{Analyses on the Grouped Parameter Sharing Mechanism}
\label{apdx:ana_group_psm}
In this section, we conduct the subjective evaluation to compare the audio quality with different numbers of shared groups ($N_g$) for PortaSpeech (normal) and PortaSpeech (small). The results are shown in \autoref{tab:ana_num_group}. It can be seen that the audio quality drops significantly when sharing parameters among all flow steps, demonstrating the effectiveness of our grouped parameter sharing mechanism.

\begin{table}[!h]
\small
\vspace{-3mm}
\caption{The audio quality (MOS-Q) and number of model parameters (\#Params.) comparisons with different number of shared groups ($N_g$). The evaluation is conducted on a server with 1 NVIDIA 2080Ti GPU and batch size 1. The mel-spectrograms are converted to waveforms using Hifi-GAN (V1)~\cite{kong2020hifi}.}
\vspace{2mm}
\centering
\begin{tabular}{ l | c | c | c }
\toprule
Method & $N_g$ &  MOS-Q & \#Params. \\
\midrule
\textit{GT}          & / & 4.43 $\pm$ 0.06 & /   \\
\textit{GT (voc.)}   & / & 4.12 $\pm$ 0.07 & /   \\
\midrule   
\multirow{4}{*}{\textit{PortaSpeech (normal)}}
& 1  & 3.86 $\pm$ 0.06 & 19.4M \\
& 3  & 3.91 $\pm$ 0.05 & 21.8M \\
& 6  & 3.93 $\pm$ 0.07 & 23.7M \\
& 12 & 3.92 $\pm$ 0.05 & 28.8M \\
\midrule
\multirow{4}{*}{\textit{PortaSpeech (small)}}
& 1  & 3.77 $\pm$ 0.06 & 6.4M \\
& 2  & 3.87 $\pm$ 0.08 & 6.7M \\
& 4  & 3.86 $\pm$ 0.05 & 7.5M \\
& 8  & 3.89 $\pm$ 0.06 & 9.0M \\

\bottomrule
\end{tabular}
\label{tab:ana_num_group}
\end{table}

\section{Potential Negative Societal Impacts} 
PortaSpeech lowers the requirements for speech synthesis service deployment (memory and CPU performance) and synthesizes high-quality speech voice, which may cause unemployment for people with related occupations such as broadcaster and radio host. In addition, there is the potential for harm from non-consensual voice cloning or the generation of fake media and the voices of the speakers in the recordings might be overused than they expect.

\end{document}